\documentclass[a4paper,12pt]{article}
 
 \usepackage[english]{babel}  
 \usepackage[utf8]{inputenc}
 \usepackage{ulem}
 \usepackage{xcolor}

\usepackage{hyperref}
 
\usepackage[totalwidth=16.5cm,totalheight=23.5cm]{geometry}
\usepackage[numbers,sort&compress]{natbib}
\usepackage[nottoc]{tocbibind}

\usepackage{amsmath,amssymb,amsthm,mathtools,mathrsfs}
\usepackage{bm,faktor}
  
  \newcommand{\cA}{\mathcal{A}}	
  \newcommand{\at}{\tilde{a}}

  \newcommand{\bbC}{\mathbb{C}}

  \newcommand{\Db}{\overline{D}}
    
  \newcommand{\gd}{\delta}

  \newcommand{\cE}{\mathcal{E}}
  
  \newcommand{\cF}{\mathcal{F}}
  
  \newcommand{\Ft}{\tilde{F}}

  \newcommand{\gi}{\iota}

  \newcommand{\gl}{\lambda}
  
  \newcommand{\gL}{\Lambda}

  \newcommand{\cN}{\mathcal{N}}				

  \newcommand{\gO}{\Omega}

  \newcommand{\pih}{\hat{\pi}}

  \newcommand{\st}{\tilde{s}}

  \newcommand{\bbT}{\mathbb{T}}	 
  \newcommand{\cT}{\mathcal{T}}				
  \newcommand{\gt}{\tau}
  
  \newcommand{\gthat}{\hat{\tau}}			

  \newcommand{\gth}{\theta}					

  \newcommand{\xih}{\hat{\xi}}

\newcommand{\from}{\colon}
\newcommand{\xto}[2][]{\xrightarrow[#1]{#2}}

\DeclareMathOperator{\SU}{SU\!}
\newcommand{\su}{\mathfrak{su}}

\DeclareMathOperator{\GL}{GL\!}
\DeclareMathOperator{\fkgl}{\mathfrak{gl}\!}

\newcommand{\parD}{\partial}
\newcommand{\parDb}{\bar{\partial}}
\newcommand{\parDh}{\hat{\partial}}
\newcommand{\LieD}{\mathcal{L}}
\newcommand{\intD}{\lrcorner}

\newcommand{\W}{\wedge}
\newcommand{\So}[1]{\Gamma\left[ #1 \right]}

\renewcommand{\S}{\bm{\text{S}}}				

\newtheorem{Proposition}{Proposition}[section]
\newtheorem{Theorem}{Theorem}[section]

\theoremstyle{definition}

\numberwithin{equation}{section} 

    \newtheorem{manualtheoreminner}{Theorem}
\newenvironment{manualtheorem}[1]{%
  \renewcommand\themanualtheoreminner{#1}%
  \manualtheoreminner
}{\endmanualtheoreminner}

\usepackage{comment}

\begin{document} 
	\pagestyle{plain}
	\title{Higher-Spin Self-Dual Yang-Mills and Gravity \\ 
	from the twistor space}
	\author{
	Yannick Herfray${}^a$, Kirill Krasnov${}^b$ and 
	Evgeny Skvortsov\footnote{Research Associate of the Fund for Scientific Research -- FNRS, Belgium}$\,{}^c$\footnote{Also on leave from Lebedev Institute of Physics, Moscow, Russia}
	\\ {}\\
	{\it ${}^a$ Institut Denis Poisson UMR 7013, Universit\'e de Tours,} \\ {\it Parc de Grandmont, 37200 Tours, France}
\\{} \\
{\it ${}^b$ School of Mathematical Sciences, University of Nottingham,} \\{\it Nottingham, NG7 2RD, UK}\\{} \\
{\it ${}^c$ Service de Physique de l’Univers, Champs et Gravitation,} \\ {\it Universit\'e de Mons, 20 place du Parc, 7000 Mons, Belgium}}
	\date{\today}
	\maketitle
\begin{abstract} We lift the recently proposed theories of higher-spin self-dual Yang-Mills (SDYM) and gravity (SDGR) to the twistor space. We find that the most natural room for their twistor formulation is not in the projective, but in the full twistor space, which is the total space of the spinor bundle over the 4-dimensional manifold. In the case of higher-spin extension of the SDYM we prove an analogue of the Ward theorem, and show that there is a one-to-one correspondence between the solutions of the field equations and holomorphic vector bundles over the twistor space. In the case of the higher-spin extension of SDGR we show show that there is a one-to-one correspondence between solutions of the field equations and Ehresmann connections on the twistor space whose horizontal distributions are Poisson, and whose curvature is decomposable. These data then define an almost complex structure on the twistor space that is integrable.  
\end{abstract}
\maketitle
\section{Introduction}
Despite the abundance of no-go theorems against theories with massless higher-spin fields (Higher spin gravities or HiSGRA) there are several examples that bypass these pitfalls: $3d$ models with topological (partially-)massless and conformal higher-spin fields \cite{Blencowe:1988gj,Bergshoeff:1989ns,Campoleoni:2010zq,Henneaux:2010xg,Pope:1989vj,Fradkin:1989xt,Grigoriev:2019xmp,Grigoriev:2020lzu}; Chiral HiSGRA \cite{Metsaev:1991mt,Metsaev:1991nb,Ponomarev:2016lrm,Skvortsov:2018jea,Skvortsov:2020wtf}; higher-spin extensions of $4d$ conformal gravity \cite{Segal:2002gd,Tseytlin:2002gz,Bekaert:2010ky}; IKKT-based theories  \cite{Sperling:2017dts,Tran:2021ukl,Steinacker:2022jjv}; collective dipole \cite{deMelloKoch:2018ivk,Aharony:2020omh}.\footnote{The latter two are strictly speaking not local field theories. Nevertheless, they come equipped with well-defined prescriptions to compute e.g. holographic correlators.} Chiral HiSGRA can be of interest as the only perturbatively local field theory with propagating massless fields, which makes it an interesting playground where the usual QFT and AdS/CFT methods apply. This is especially in view of the recent developments such as one-loop UV-finiteness \cite{Skvortsov:2018jea,Skvortsov:2020wtf,Skvortsov:2020gpn} as well as fully covariant equations of motion of Chiral HiSGRA  \cite{Skvortsov:2022syz,Sharapov:2022faa,Sharapov:2022awp,Sharapov:2022wpz}. However, as is customary for self-dual theories, the interactions are complex in Minkowski signature. Nevertheless, Chiral HiSGRA should be a consistent truncation of the holographic dual of Chern-Simons matter theories \cite{Sharapov:2022awp}. Therefore, all solutions and amplitudes of Chiral theory should carry over to the complete theory.

Chiral HiSGRA admits two simple consistent truncations where either Yang-Mills or gravitational interactions are retained and the scalar field is dropped, which was shown in \cite{Ponomarev:2017nrr} in the light-cone gauge. These two truncations can be thought of as higher-spin extensions of the self-dual Yang-Mills (SDYM) and self-dual Gravity (SDGR) theories, HS-SDYM and HS-SDGR, respectively. In \cite{Krasnov:2021nsq} two of the present authors have constructed the covariant formulations of these HS-SDYM and HS-SDGR. The question left unanswered was whether there exist a twistor space description of these theories. An attempt in this direction using the standard projective twistor space was made in \cite{Tran:2021ukl}. Another attempt at a twistor description of higher spins is a recent paper \cite{Tran:2022tft}, where a twistor description of the full chiral higher-spin gravity is contemplated. For applications of twistor techniques to conformal HiSGRA see \cite{Hahnel:2016ihf,Adamo:2016ple}.

This work is a natural continuation of \cite{Krasnov:2021nsq} in that it gives an answer to the question of the twistor space description of self-dual higher-spin theories. Our main new insight is that the appropriate room for the higher-spin (self-dual) theories is not in the projective twistor space, but rather in the full spinor bundle over the four-dimensional manifold in question. The usual SDYM and SDGR then arise as the subsectors of the higher-spin extensions that descend to the projective twistor space. 

In the case of HS-SDYM we provide both a description of the twistor space lift of the theory, as well as a statement in the opposite direction, i.e. an analog of the Ward theorem. Our main statement can be formulated as the following theorem.

Let $S' \xto{\pi} \S^4$ be the bundle of (primed) spinors on $S^4$, the twistor space $\bbT = (\bbC^4)^* \xto{\pi} \S^4 $ is obtained by deleting from $S'$ the zero section.  Let $U$ be an open set of $S^4$ and let $V = \pi^*(U)$ be the corresponding open set of $\bbT$.
\begin{manualtheorem}{A}
	There is a one-to-one correspondence between solutions of the higher-spin self-dual Yang-Mills equations on $U \subset S^4$ (up to a gauge) and holomorphic bundles $\cE \to V$ such that the restriction of $\cE$ along each of the fibres of $V \to U$ is holomorphically trivial.
\end{manualtheorem}
This theorem can be viewed as the higher-spin analogue of the Ward theorem. 

\bigskip
In the case of HS-SDGR we have the following statement. Let $S' \xto{\pi} M^4$ be the bundle of (primed) 2-component spinors on a 4-manifold $M^4$, the twistor space $\cT \xto{\pi} M^4 $ is obtained by deleting from $S'$ the zero section.  Let $\mathcal{V} \subset T\cT$ be the vertical distribution. Let $U$ be an open set of $M^4$ and let $V = \pi^*(U)$ be the corresponding open set of $\cT$. We define the horizontal distributions on $TV$ to be those in the kernel of the projection $P \from T\cT \to \mathcal{V}$
	\begin{equation*}
	P =  \gt^{A'}\frac{\parD}{\parD \pi^{A'}} +  \gthat^{A'}\frac{\parD}{\parD \pih^{A'}}, \qquad \tau^{A'}:= d\pi^{A'} + \cA^{A'}(x,\pi,\hat{\pi})\,.
	\end{equation*}
	These are parametrised by the Ehresmann connection $\cA^{A'}(x,\pi,\hat{\pi})$. Here $A'$ is the 2-component spinor index, $\pi^{A'}$ is the fibre coordinate and $\hat{\pi}^{A'}$ is constructed by $\pi^{A'}$ using the hat operation that is available in the Euclidean signature and squares to minus the identity (see section \ref{sec:SDYM} for more details). The connection  $\cA^{A'}(x,\pi,\hat{\pi})$ is a one-form on $M^4$. The twistor space naturally is a Poisson manifold with Poisson structure
	\begin{equation}\label{Intro: Poisson structure}
	  \epsilon^{B'A'} \frac{\parD}{\parD \pi^{A'}}  \frac{\parD}{\parD \pi^{B'}} +  \epsilon^{B'A'} \frac{\parD}{\parD \pih^{A'}}  \frac{\parD}{\parD \pih^{B'}}\,,
	  \end{equation}
	where $\epsilon^{A'B'}$ is the inverse of the volume form on the $(\mathbb{C}^2)^*$ fibres.
	
\begin{manualtheorem}{B}
For horizontal distributions 
that are infinitesimal symmetries of the Poisson structure \eqref{Intro: Poisson structure} the Ehresmann connection $\cA^{A'}$ has a potential: $\cA^{A'}=-\epsilon^{A'B'} \partial_{B'} A$, where $A=A(x,\pi)$. In particular, the Ehresmann connection $\cA^{A'}$ of Poisson horizontal distributions is independent of $\hat{\pi}$. Furthermore, its curvature 2-form has also a potential $\cF^{A'}=-\epsilon^{A'B'}\partial_{B'} F$, where $F=F(x,\pi)=dA+(1/2)\{A,A\}$ and $\{\cdot,\cdot\}$ is the Poisson bracket given by \eqref{Intro: Poisson structure}. There is a one-to-one correspondence (up to a gauge) between solutions of the higher-spin self-dual gravity equations on $U$ and Poisson horizontal distributions on $TV$ whose curvature potential $F$ is  decomposable $F\W F=0$. What is more, the two simple factors of $F$ define, together with the 1-forms $\tau^{A'}$, an almost complex structure on $V$ that is integrable. 
\end{manualtheorem}

In both the cases of HS-SDYM and HS-SDGR we explain the geometric origin of the subtle invariances that the field equations possess. These are seen to come from the diffeomorphisms of the twistor space.

The organisation of this paper is very simple: in sections \ref{sec:SDYM} and \ref{sec:SDGR} we discuss (higher-spin extensions) of self-dual Yang-Mills theory and self-dual gravity, respectively. 

\section{Higher-spin self-dual Yang-Mills in twistor space}
\label{sec:SDYM}
Let $P \to S^4$ be a $GL(N,\bbC)$-principal bundle and let $E \to S^4$ be some associated bundle.

\subsection{HS-SDYM equations: Spacetime equations}
We follow \cite{Krasnov:2021nsq} in this subsection. The ``Higher-spin YM potential'' is a (formal) sum of Lie algebra-valued one-forms with different numbers of primed spinor indices
\begin{equation}\label{HS-SD_YM Spacetime: Potential}
A = \sum_{s=1}^{\infty} A{}^{A'(2s-2)} = A+ A^{A'(2)}+ A^{A'(4)}+\ldots
\end{equation}
where all fields take values in the Lie algebra of $\GL\left(N,\bbC\right)$ and are also one-forms. The first term in the sum on the right-hand side is the usual Yang-Mills gauge potential -- a Lie algebra-valued one-form. The notation $A'(n)$ is standard in the higher-spin literature, and denotes $n$ different primed indices that are symmetrized. Thus, 
$$ A'(n) \equiv ( A'_1 \ldots A'_n)\,.$$
For any 
\begin{equation}
    \xi = \sum_{s=1}^{\infty} \xi^{A'(2s-2)} = \xi + \xi^{A'(2)}+\xi^{A'(4)}+\ldots
\end{equation}
taking values in $\fkgl\left(N,\bbC\right)$, the ``higher-spin gauge transformations'' are
\begin{equation}\label{HS-SD_YM Spacetime: gauge transformation}
\gd_{\xi} A = \sum_{s=1}^{\infty} d\xi^{A'(2s-2)} + \sum_{s=1}^{\infty}\sum_{\st=1}^{\infty} ([A, \xi])^{A'(2s+2\st-4)}\,.
\end{equation}
It is useful to write the first couple of terms in this series explicitly 
\begin{eqnarray}
    \delta_\xi A = (d\xi +[A,\xi])+ ( d\xi^{A'(2)}+ [A,\xi^{A'(2)}]+[A^{A'(2)},\xi])+\\ \nonumber
    ( d\xi^{A'(4)}+ [A,\xi^{A'(4)}]+[A^{A'(4)},\xi]+[A^{A'(2)},\xi^{A'(2)}])+\ldots
\end{eqnarray}
The first term in this sum is the usual gauge transformation of Yang-Mills field. It is important to note that in the last term in the second line the spinor indices of $A^{A'(2)}$ and $\xi^{A'(2)}$ are assumed to be symmetrized, so that only the terms with totally symmetric spinor indices arise. This can easily be implemented by defining generating functions 
\begin{align*}
   A(\pi)&= \sum_{s=1}^{\infty} A{}^{A'(2s-2)}\, \pi_{A'}\ldots \pi_{A'}\,,  &&\xi(\pi)= \sum_{s=1}^{\infty} \xi^{A'(2s-2)}\pi_{A'}\,\ldots \pi_{A'}
\end{align*}
with the help of an auxiliary commuting variable $\pi_{A'}$. 

The corresponding ``field strength'' is defined as
\begin{align}\label{HS-SD_YM Spacetime: Field strength}
 F = \sum_{s=1}^{\infty} dA^{A'(2s-2)} + \sum_{s=1}^{\infty}\sum_{\st=1}^{\infty} \frac{1}{2}\left[ A \W A \right] ^{A'(2s+2\st-4)}\,.
\end{align}
Again, the spinor indices in a product expression are always symmetrized to produce only the terms with totally symmetric spinor indices. Restricting to the first terms in the above sum we recover the usual Yang-Mills curvature.

The higher-spin self-dual Yang-Mills equations are
\begin{equation}\label{ASD-projection}
F\big|_{ASD} = 0\,
\end{equation}
where ``ASD'' stands means that the curvature is restricted to its anti self dual part. To give this more concrete meaning we recall that the anti self dual (ASD in what follows) projection of a 2-form is computed by converting the spacetime indices into the spinor ones $\mu\to MM'$. A 2-form $B$ is then split into its SD and ASD parts as follows
\begin{eqnarray}
    B_{MM'NN'} = \frac{1}{2}B_{(ME'N)}{}^{E'} \epsilon_{M'N'}+ \frac{1}{2}B_{E(M'}{}^{E}{}_{N')} \epsilon_{MN}\,.
\end{eqnarray}
(By convention the primed indices cannot be mixed with the unprimed ones. This has the effect that their relative positions are irrelevant. For example, it does not matter whether or not the parenthesis denoting the symmetrization encompass a single primed index).  The second terms is the ASD part of the 2-form $B$. Thus, in (\ref{ASD-projection}) there is a pair of primed spinor indices coming from the projection of the 2-form onto its ASD part, as well as the primed spinor indices that are labels of the different summands in $F$. It is assumed that all the primed spinor indices are symmetrized in (\ref{ASD-projection}). Let us write down the first few equations contained in (\ref{ASD-projection}) explicitly. Thus, 
\begin{equation}
    d_{E}{}^{(A'_1} A^{EA'_2)} + \frac{1}{2} [A_E{}^{(A'_1}, A^{E A'_2)}]=0
\end{equation}
is the usual SDYM equation. The next equation has four free primed spinor indices and reads
\begin{equation}
    d_{E}{}^{(A'_1} A^{EA'_2 A'_3 A'_4)} +  [A_E{}^{(A'_1}, A^{E A'_2 A'_3 A'_4)}]=0\,.
\end{equation}

Because the spinor index that comes from the one-form index is always symmetrized with the other primed spinor indices in the above equations, the arising equations enjoy an extra gauge invariance under shift symmetry:
\begin{equation}\label{HS-SD_YM Spacetime: shift invariance}
\gd A = dx^{BB'} \; \sum_{s=2}^{\infty} \gd_{B'}{}^{A'} \eta_{B}{}^{A'(2s-4)} = dx^{B(A'_1} \eta_B^{A'_2)}+\ldots
\end{equation}
As opposed to the ``higher-spin gauge transformations'', which reduce to usual gauge transformations when restricted to the $s=1$ terms, the shift invariance is a genuine new feature of the higher-spin equations and is another type of gauge symmetry that the equations enjoy.

\subsection{Geometrical realisation in twistor space}
Let $S' \xto{\pi} \S^4$ be the bundle of (primed) spinors on $S^4$. We will denote the primed spinor that is a coordinate along the fibre by $\pi_{A'}$. The twistor space $\bbT = (\bbC^4)^*$ is obtained by deleting from $S'$ the zero section. The Euclidean spinors admit a hat operator. This is an anti-linear operator that maps primed spinors to primed spinors and squares to minus the identity. The existence of this operator follows from the fact that Euclidean spinors are $\SU(2)$ spinors and therefore posses both an invariant hermitian inner product $h_{A'\bar{A}'}$ and an invariant volume form $\epsilon_{A'B'}$. The hat operator is then simply obtained by
\begin{equation}\label{hat operator}
\pi^{A'} \mapsto \pih^{A'} := \epsilon^{A'B'}h_
{B'\bar{B}'} (\pi^*)^{\bar{B}'},
\end{equation}
(where the star indicates the usual complex conjugation). We will make extensive use of the following identities:
\begin{equation}\label{pi-identity}
\pi_{B'}\pih^{A'} - \pih_{B'} \pi^{A'} = \langle \pi \hat{\pi}\rangle \; \gd_{B'}{}^{A'}\,,
\end{equation}
where we introduced 
\begin{eqnarray}
    \left( \pi_{D'}\pih^{D'} \right):=\langle \pi \hat{\pi}\rangle\,.
\end{eqnarray}
The complex structure on $\bbT$ that we use is as follows. The basis of $(0,1)$ 1-forms is given by
\begin{eqnarray}
    d\hat{\pi}^{A'}, \qquad \hat{\pi}_{A'} dx^{AA'}\,,
\end{eqnarray}
with dual vector fields
\begin{eqnarray}
    \frac{\partial}{\partial \hat{\pi}^{A'}}\,, \qquad -\frac{\pi^{A'}\partial_{AA'}}{\langle \pi \hat{\pi}\rangle}\,.
\end{eqnarray}
Altogether the corresponding Dolbeault operator is
\begin{equation}\label{Dolbeault operator}
    \parDb = d\hat{\pi}^{A'}\frac{\partial}{\partial \hat{\pi}^{A'}} -\hat{\pi}_{C'} dx^{AC'}\frac{\pi^{B'}\partial_{AB'}}{\langle \pi \hat{\pi}\rangle}.
\end{equation}
Making use of the identity \eqref{pi-identity}, the projection of $A_{AA'}dx^{AA'}$ on its $(1,0)$ and $(0,1)$ parts are
\begin{align*}
A_{AA'} dx^{AA'} \big|_{1,0} &=  \frac{ A_{AA'}\pih^{A'}}{\langle \pi \hat{\pi}\rangle } \;dx^{AB'}\pi_{B'}, &
A_{AA'} dx^{AA'} \big|_{0,1} &=  - \frac{A_{AA'}\pi^{A'}}{\langle \pi \hat{\pi}\rangle } \; dx^{AB'}\pih_{B'}\,.
\end{align*}

\subsubsection{Higher-Spin Yang-Mills fields on twistor space}
The higher-spin Yang-Mills potential \eqref{HS-SD_YM Spacetime: Potential} that was introduced as a formal sum of Lie algebra-valued one-forms with a different number of spinor indices naturally arises from the following field on the twistor space,
\begin{equation}\label{HS-SD YM Spacetime: twistor gauge potential1}
\cA(x,\pi) := \sum_{s=1}^{\infty} A(x){}^{A'(2s-2)} \pi_{A'(2s-2)}.
\end{equation}
This reinterprets the connection as a gauge potential in the
pullback bundle $\pi^*E \to \bbT$, which is a gauge bundle where the base is now the full twistor space $\bbT$, and not just its projectivised version as in the standard construction. One can then readily see that, modulo the terms involving $d\pi_{A'}$, the higher-spin gauge transformations \eqref{HS-SD_YM Spacetime: gauge transformation} coincide with the usual gauge transformations in the  twistor space
\begin{equation*}
\gd_{\xi} \cA = d \xi + \left[\cA, \xi\right]\,,
\end{equation*}
and the higher-spin field strength \eqref{HS-SD_YM Spacetime: Field strength} coincides with the usual curvature 2-form
\begin{equation*}
\cF = d \cA + \frac{1}{2}[\cA , \cA]\,,
\end{equation*}
again modulo the terms involving $d\pi_{A'}$.
In turn, the HS-SDYM field equations are clearly related to 
\begin{equation*}
\cF \big|_{0,2}= 0\,,
\end{equation*}
again modulo the terms along the fibre direction. We now develop a holomorphic bundle interpretation that clarifies all these issues. 
\subsubsection{Holomorphic bundle interpretation}
 Just as in the usual Ward correspondence one can introduce the $(0,1)$ part of the gauge field
\begin{align}\label{HS-SD YM Spacetime: twistor gauge potential2}
a := -\sum_{s=1}^{\infty} \left(A{}^{A'(2s-2)} \pi_{A'(2s-2)} \right)\Big|_{0,1}
 = \frac{dx^{AB'}\pih_{B'}}{\langle \pi \hat{\pi}\rangle}\; \sum_{s=1}^{\infty}  A_{A}{}^{A'(2s-1)} \pi_{A'(2s-1)}\,.
\end{align}
Note that $2s-2$ of the primed indices on $A_A{}^{A'(2s-1)}$ are those that existed already in (\ref{HS-SD_YM Spacetime: Potential}), while an additional primed index arises when the one-form index of the gauge potential is converted into a pair of unprimed and primed spinors. 

Note that the projection onto the part of the higher-spin gauge potential (\ref{HS-SD_YM Spacetime: Potential}) that is invariant under (\ref{HS-SD_YM Spacetime: shift invariance}) has already occurred here, because the primed spinor index that came from the one-form index became symmetrized with the other primed indices of the gauge potentials. Indeed, under the shift symmetry \eqref{HS-SD_YM Spacetime: shift invariance} the higher-spin potential \eqref{HS-SD YM Spacetime: twistor gauge potential1} is shifted by a $(1,0)$ term and therefore \eqref{HS-SD YM Spacetime: twistor gauge potential2} is invariant. 

Recall that $A$ is a connection on the associated bundle $E \to S^4$. Then \eqref{HS-SD YM Spacetime: twistor gauge potential2} defines a differential operator on the pull-back bundle $\pi^*E\to \bbT$ via
	\begin{equation*}
\Db \from \left| 
\begin{array}{ccccc}
\So{\pi^*E}& \to & \gO^{0,1}\left(\bbT, \pi^*E\right) \\
\Phi & \mapsto & \left( \parDb + a \right) \Phi\,
\end{array}\right. ,
\end{equation*} 
which maps sections of the bundle $\pi^* E$ into $(0,1)$ forms on $\bbT$ valued in the space of sections.  The HS-SDYM field equations are then the conditions for $\pi^*E \to \bbT$ to be holomorphic:
\begin{equation}\label{HS-SD YM Spacetime: field equations2}
\parDb a + \frac{1}{2}[a , a] = 0\,.
\end{equation}
Seeing this is an exercise. Some key steps of this computation, which illustrate how it works, are as follows. First, we have
\begin{eqnarray}
    d\left( \frac{dx^{AB'} \hat{\pi}_{B'}}{\langle \pi \hat{\pi}\rangle}\right)= - \frac{dx^{AB'}\wedge d\hat{\pi}_{B'}}{\langle \pi \hat{\pi}\rangle} - d\langle \pi \hat{\pi}\rangle \wedge \frac{dx^{AB'} \hat{\pi}_{B'}}{\langle \pi \hat{\pi}\rangle^2}\,.
\end{eqnarray}
We need to project it onto the $(0,2)$ part. This is done by inserting the identity (\ref{pi-identity}) in the first term, and then keeping only the $\pi d\hat{\pi}$ part in the second term. This gives
\begin{eqnarray}
    d\left( \frac{dx^{AB'} \hat{\pi}_{B'}}{\langle \pi \hat{\pi}\rangle}\right)\big|_{0,2}=  \frac{dx^{AB'}\hat{\pi}_{B'} \wedge \pi^{C'} d\hat{\pi}_{C'}}{\langle \pi \hat{\pi}\rangle^2} -  \pi_{C'} d\hat{\pi}^{C'} \wedge \frac{dx^{AB'} \hat{\pi}_{B'}}{\langle \pi \hat{\pi}\rangle^2}=0\,.
\end{eqnarray}
Thus, applying the exterior derivative to the potential (\ref{HS-SD YM Spacetime: twistor gauge potential2}) and keeping the projection onto the $(0,2)$ part gives
\begin{eqnarray}
    \frac{\Sigma^{B'C'}\pih_{B'}\pih_{C'}}{\langle \pi \hat{\pi}\rangle^2}  \sum_{s=1}^{\infty}  \partial_{A}{}^{D'} A^{A A'(2s-1)} \pi_{A'(2s-1)}\pi_{D'}
\end{eqnarray}
where we used
\begin{eqnarray}
    (dx^{AB'}\pih_{B'})\wedge(dx^{BC'}\pih_{C'}) = \epsilon^{AB} \Sigma^{B'C'} \pih_{B'}\pih_{C'}, \qquad \Sigma^{B'C'}:= \frac{1}{2} dx_A{}^{B'}\wedge dx^{AC'}.
\end{eqnarray}
Similarly, taking the $[a,a]$ we get
\begin{eqnarray}
    [a,a]= \frac{\Sigma^{B'C'}\pih_{B'}\pih_{C'}}{\langle \pi \hat{\pi}\rangle^2} \sum_{s=1}^{\infty}  \sum_{\tilde{s}=1}^\infty [A_{A}{}^{A'(2s-1)}, A^{A}{}^{A'(2\tilde{s}-1)}] \pi_{A'(2s-1)} \pi_{A'(2\tilde{s}-1)}\,.
\end{eqnarray}
Making use of this and of the explicit form \eqref{Dolbeault operator} of the Dolbeault operator, one get that \eqref{HS-SD YM Spacetime: field equations2} coincides with the higher-spin SDYM equations
\begin{eqnarray}
    \partial_{A}{}^{A'} A^{A A'(2s-1)}+ \frac{1}{2} \sum_{s=1}^{\infty}  \sum_{\tilde{s}=1}^\infty [A_{A}{}^{A'(2s-1)}, A^{A}{}^{A'(2\tilde{s}-1)}] =0\,.
\end{eqnarray}

Let us make a final important remark: while \eqref{HS-SD YM Spacetime: twistor gauge potential1} clearly receives the interpretation of a connection on the whole of $\pi^* E \to S'$ (not just on $\pi^* E \to \bbT$), one however sees from \eqref{HS-SD YM Spacetime: twistor gauge potential2} that, for every fixed $x$, $a(x, \gl \pi)$ does not have a well defined limit as $|\gl| \to 0$ and thus does not continuously extend on $S'$. This stems from the fact that there is no complex structure defined along the zero section of $S' \to S^4$: the projection
\begin{align*}
A_{AA'} dx^{AA'} \big|_{0,1} &=  - \frac{A_{AA'}\pi^{A'}}{\pi_{D'}\pih^{D'} } \; dx^{AB'}\pih_{B'}.
\end{align*}
can only  be continuous at $\pi=0$ if $A_{AA'}=0$. So, it is important to remove the zero section from the total bundle of spinors to be able to give the twistor interpretation to the whole construction. 

\subsection{A higher-spin Ward correspondence}

\subsubsection{Correspondence}
Let $S' \xto{\pi} \S^4$ be the bundle of spinors on $S^4$, the twistor space $\bbT = (\bbC^4)^* \xto{\pi} \S^4 $ is obtained by deleting from $S'$ the zero section.  Let $U$ be an open set of $S^4$ and let $V = \pi^*(U)$ be the corresponding open set of $\bbT$.
\begin{Theorem}
	There is a one-to-one correspondence between solutions of the higher-spin self-dual Yang-Mills equations on $U \subset S^4$ (up to a gauge) and holomorphic bundles $\cE \to V$ such that the restriction of $\cE$ along each of the fibres of $V \to U$ is holomorphically trivial.
\end{Theorem}
 This theorem should be understood as a "higher-spin Ward theorem", generalizing the classical work \cite{Ward:1977ta} to higher spins: Holomorphic bundles on twistor space $\bbT$ correspond to higher-spin self-dual Yang-Mills solutions. The geometrical restriction to bundles which descend to {\it projective} twistor space $\textrm{P}\bbT$ amounts to restricting the higher-spin fields to the usual self-dual Yang-Mills.
\subsubsection{Proof}
	Let $\cE \to V$ be a holomorphic bundle satisfying the requirements of the theorem. 
	 It can equivalently be represented by a differential operator
	\begin{equation*}
	\Db \from \left| 
	\begin{array}{ccccc}
	 \So{\cE}& \to & \gO^{0,1}\left[\cE\right] \\
	 \Phi & \mapsto & \left( \parDb + a \right) \Phi
	\end{array}\right.
	\end{equation*} 
with vanishing ``curvature''
\begin{align*}
\Db^2  & = \parDb a + \tfrac{1}{2}[a,a]  = 0\,.
\end{align*}
In a local patch, we can write
\begin{equation*}
a = a_{A}\; dx^{AB'}\pih_{B'} + \at_{A'} \;d\pih^{A'}\,,
\end{equation*}
and the curvature then is
\begin{align} \label{HS-SD_YM Twistorspace: Ward transform proof_curvature}
\Db^2  & = \left( \frac{\parD}{\parD \pih^{A'}}\at_{B'} +  \tfrac{1}{2}[\at_{A'},\at_{B'}]  \right)  d\pih^{A'} \W d\pih^{B'}\\
&+ \left( \frac{\parD}{\parD \pih^{A'}}a_{A} +\frac{ \pi_{A'}}{\langle \pi \hat{\pi}\rangle} a_{A} + \frac{\pi^{C'}}{\langle \pi \hat{\pi}\rangle}\partial_{AC'}\at_{A'}  + \left[  \at_{A'} ,  a_{A}\right]   \right)  d\pih^{A'} \W dx^{AB'}\pih_{B'}\nonumber\\
&+ \left( -\frac{\pi^{C'}}{\langle \pi \hat{\pi}\rangle} \partial_{AC'} a_{B} +  \tfrac{1}{2}[a_{A}, a_{B}]  \right)  dx^{AA'}\pih_{A'} \W dx^{BB'}\pih_{B'}\,.\nonumber
\end{align}
The vanishing of the first line means that the restriction of $\cE \to V$ along each of the $(\bbC^2)^*$ fibres is holomorphic. This is because the restriction of $\Db$ to the fibres is
\begin{equation*}
d\pih^{A'}\left(  \frac{\parD}{\parD \pi^{A'} } + \at_{A'} \right).
\end{equation*}
By hypothesis this bundle is holomorphically trivial.\footnote{We would like to thank Lionel Mason for a discussion that helped us to clarify this point.} Therefore one can choose\footnote{Note that as opposed to the usual SDYM correspondence where one can always make this choice of gauge, this here really makes use of the first of the field equations \eqref{HS-SD_YM Twistorspace: Ward transform proof_curvature}.}  a trivialisation such that $\at_{A'} =0$. The point here is that a holomorphic bundle can be topologically trivial but not holomorphically trivial. Let us see this for the simplest case of line bundles (corresponding to abelian theories on spacetime). Suppose that $E \to(\mathbb{C}^2)^*$ is topologically trivial i.e. $E = \mathbb{C} \times (\mathbb{C}^2)^*$. Then a choice of a holomorphic structure is equivalent to choosing a flat Dolbeault operator: $\bar \partial_{A'} + a_{A'}$ with $\bar \partial a =0$ up to gauge $a \sim a + \bar\partial f$. In other terms, they are given by the cohomology group $H^{0,1}((\mathbb{C}^2)^* , \mathbb{C} )$, which is non trivial. This is as opposed to the usual Ward result where one would be considering $H^{0,1}(\mathbb{C}P^1 , \mathbb{C} )=0$.

Within this assumption, and introducing $$\cA_A\left(x, \pi\right) := \langle \pi \hat{\pi}\rangle a_{A},$$ the second line of the curvature \eqref{HS-SD_YM Twistorspace: Ward transform proof_curvature} is found to be equivalent to 
$$\frac{\parD}{\parD \pih^{A'}}\cA_{A}=0\,.$$
Therefore $\cA_{A}$ is a holomorphic function along the $(\bbC^2)^*$ fibres of $V \to U$. However, by Hartogs's extension theorem there do not exist isolated singularities in complex dimension $n\geq2$ and therefore for every $x\in U$, $\cA_{A}(x,\pi)$ can be holomorphically extended to the whole of the $\bbC^2$ fibres. In particular we must have
\begin{equation*}
\cA_{A}\left(x,\pi,\pih\right) =   \sum_{s=1}^{\infty}  A_{A}{}^{A'(2s-1)}(x) \pi_{A'(2s-1)}.
\end{equation*}
Strictly speaking, we should sum over all spins in the above sum, both integers and half-integers. However, one can easily restrict the sum to integer spins only by requiring $A$ to be invariant under parity on $\mathbb{C}^2$. This reproduces the ansatz for the connection \eqref{HS-SD YM Spacetime: twistor gauge potential1}. The vanishing of the third line \eqref{HS-SD_YM Twistorspace: Ward transform proof_curvature} is then equivalent to the higher-spin self-dual Yang-Mills equations.
 $\qed$
\section{Higher-Spin Self-Dual Gravity in twistor space}
\label{sec:SDGR}
Let $P \to M^4$ be a $SU(2)$-principal bundle on $M^4$. Let $S' \xto{\pi} M$ be the bundle of spinors, defined as the associated bundle for the fundamental representation of $SU(2)$. The twistor space $\cT$ is obtained by deleting from $S'$ the zero section. 

\subsection{HS-SDGR equations: Spacetime equations}

\subsubsection{HS-SDGR equations $\gL \neq0$}
The ``Higher-spin gravity potential'' is best described using its generating functional, as in \cite{Krasnov:2021nsq}. We define
\begin{equation}\label{HS-SD_GR Spacetime: Potential}
A = \sum_{n=2}^{\infty} \frac{1}{n!} A^{A'(n)} \pi_{A'_1} \ldots \pi_{A'_n},
\end{equation}
where every term is one-form valued. It is assumed that the sum here is taken over even spins only, which is easily imposed by requiring the potential to be invariant under $\pi_{A'} \to -\pi_{A'}$.

The ``field strength'' is defined by
\begin{align}\label{HS-SD_GR Spacetime: Field strength}
F := dA + \frac{1}{2}\{ A,A\},
\end{align}
where we introduced the Poisson bracket of functions of $\pi_{A'}$
\begin{eqnarray}\label{Poisson}
   \{ f(\pi), g(\pi)\} = \partial^{C'} f \partial_{C'} g= 
   \sum_n \frac{1}{n!} \sum_{k+m=n} \frac{n!}{k! m!} f^{A'(k) C'} g^{B'(m)}{}_{C'} \pi_{A'_1} \ldots\pi_{A'_k}\pi_{B'_1} \ldots\pi_{B'_m}.
\end{eqnarray}
The higher-spin self-dual gravity equations are
\begin{equation}\label{HS-SD_GR Spacetime: Equations}
 F \wedge F  =0.
\end{equation}
This, in particular, implies that $F$ is a decomposable 2-form, fact which will be of importance below. Restricting to the first terms in the above sums we recover respectively, an $\su(2)$-valued connection $A^{A'B'}$, its curvature $F^{A'B'}$ and $\gL \neq0$ self-dual gravity equations $F^{(A'B'}\W F^{C'D')}=0$. For more information about the spin-2 case we refer the reader to \cite{Krasnov:2016emc}, see also \cite{Herfray:2016qvg} for a thorough discussion on how the field equations $F^{(A'B'}\W F^{C'D')}=0$ relate to the Mason--Wolf twistor action \cite{Mason:2007ct}. 

\subsubsection{Gauge invariance}
These equations enjoy several gauge symmetries. First, for any
\begin{equation*}
\xi =  \sum_{n=2}^{\infty} \frac{1}{n!} \xi^{A'(n)}\pi_{A'_1} \ldots \pi_{A'_n}
\end{equation*}
we have the ``higher-spin gauge transformations''
\begin{equation}\label{HS-SD_GR Spacetime: gauge transformations}
\gd_{\xi} A = d\xi +\{ A,\xi\}
\end{equation} 
and for any
\begin{equation*}
\eta^{\mu} =  \sum_{n=0}^{\infty}\frac{1}{n!} \eta^{\mu}{}^{A'(n)}\pi_{A'_1} \ldots \pi_{A'_n}
\end{equation*}
the ``higher-spin generalised diffeomorphisms''
\begin{equation}\label{HS-SD_GR Spacetime: generalised diffeomorphisms}
\gd_{\eta} A = dx^{\nu} \; \eta^\mu F_{\mu\nu}\,.
\end{equation}
The first terms in these sums (i.e. the terms corresponding to the spin two case) respectively give standard $\su(2)$ gauge transformations $\gd_{\xi}A = d_{A}\xi$ and Lie derivatives (up to a gauge transformation): $\gd_{\eta}A =  \gi_\eta F=  \LieD_{\eta} A - d_{A}\left(\gi_\eta A\right)$.

\subsection{Geometrical realisation in twistor space}

\subsubsection{Higher-spin fields as a connection on the twistor space}
Let $\left(x, \pi^{A'}\right)$ be local coordinates on $\cT \to M$. It will here be crucial that the fibers are equipped with a preferred volume form $\epsilon_{A'B'}$ and that the twistor space is therefore equipped with the Poisson structure
\begin{equation}\label{HS-SD_GR Twistorspace: Poisson structure}
    \epsilon^{B'A'}\frac{\parD}{\parD \pi^{A'}}\frac{\parD}{\parD \pi^{B'}}
    + \epsilon^{B'A'}\frac{\parD}{\parD \hat{\pi}^{A'}}\frac{\parD}{\parD \hat{\pi}^{B'}}.
\end{equation}
We will start by considering the following 1-form on the twistor space
\begin{equation}\label{HS-SD_GR Twistorspace: Potential}
\gt^{A'}\parD_{A'} = \left(d\pi^{A'} + \cA(x,\pi,\hat{\pi})^{A'} \right)\frac{\parD}{\parD \pi^{A'}}\,.
\end{equation}
This object receives the following geometric interpretation: at every point $p\in \cT$ 
\begin{equation*}
P = \gt^{A'}\frac{\parD}{\parD \pi_{A'}} + \gthat^{A'}\frac{\parD}{\parD \pih_{A'}}
\end{equation*}
defines a projector $P_{p} \from T_p\cT \to \mathcal{V}_p$ on the vertical tangent bundle. This splits the tangent bundle as
\begin{equation*}
T_p \cT = \mathcal{V}_p + H_p
\end{equation*}
where $H_p$ is the kernel of $P_p$. The corresponding ``horizontal'' distribution $H$ on $TS' \to M$ is a connection in the sense of Ehresmann. A general horizontal vector field is one of the form $\xi = V^{\mu}(x,\pi,\pih) D_{\mu}$ with
\begin{equation}\label{HS-SD_GR Twistorspace: Horizontal vector field}
 D_{\mu}= \parD_{\mu} - \cA^{A'}_{\mu}(x,\pi,\pih)\parD_{A'} - \hat{\cA}^{A'}_{\mu}(x,\pi,\pih)\parDh_{A'}.
\end{equation}
We have the following proposition:
\begin{Proposition}\label{prop-horizontal}
A general horizontal vector field $\xi$ is an infinitesimal symmetry of the Poisson structure
\begin{equation}
\LieD_{\xi}\left( \epsilon^{A'B'}\frac{\parD}{\parD \pi^{A'}}\frac{\parD}{\parD \pi^{B'}}
    + \epsilon^{A'B'}\frac{\parD}{\parD \hat{\pi}^{A'}}\frac{\parD}{\parD \hat{\pi}^{B'}} \right) = 0 
\end{equation}
if and only if $\parD_{A'}V^{\mu} = \parDh_{A'}V^{\mu} = 0$ and
\begin{align}
\parDh_{B'} \cA^{A'}{}_{\mu} &=0,  & \parD_{[B'} \cA_{A']}{}_{\mu} &=0.
\end{align}
(together with the corresponding equations for the ``hatted'' connection field). These can always be solved locally as 
\begin{equation}\label{HS-SD_GR Twistorspace: Potential as HS field}
\cA(x,\pi)^{A'} :=-\epsilon^{A'B'}\partial_{B'} A(x,\pi),
\end{equation}
for some $A(x,\pi)$.
\end{Proposition}

Proof is by a computation. We have
\begin{equation}
    \LieD_{\xi}\left( \epsilon^{A'B'}\frac{\parD}{\parD \pi^{A'}}\frac{\parD}{\parD \pi^{B'}}\right)= \epsilon^{A'B'} [\xi,\frac{\parD}{\parD \pi^{A'}} ]\otimes \frac{\parD}{\parD \pi^{B'}}+ \epsilon^{A'B'}
    \frac{\parD}{\parD \pi^{A'}}\otimes [\xi, \frac{\parD}{\parD \pi^{B'}}] ,
\end{equation}
and
\begin{equation}
    [\xi, \frac{\parD}{\parD \pi^{A'}}]=
    - (\partial_{A'}V^\mu)(\partial_\mu - \cA^{C'}_\mu \partial_{C'} - \hat{\cA}^{C'}_\mu \hat{\partial}_{C'}) + V^\mu( \partial_{A'} \cA^{C'}_\mu \partial_{C'} + \partial_{A'}\hat{\cA}^{C'}_\mu
\hat{\partial}_{C'}).
\end{equation}
The absence of the $\partial_\mu \otimes \partial_{A'}$ terms in the Lie derivative implies $\partial_{A'} V^\mu=0$, and the similar reasoning for the $\partial_\mu \otimes \hat{\partial}_{A'}$ terms implies $\hat{\partial}_{A'} V^\mu=0$. On the other hand, the $\partial_{A'}\otimes\partial_{B'}$ terms in the Lie derivative are
\begin{equation}
   \epsilon^{B'A'} V^\mu \partial_{A'}\cA^{C'}_\mu \partial_{C'}\otimes \partial_{B'} + \epsilon^{B'A'} \partial_{A'} \otimes V^\mu \partial_{B'} \cA^{C'}_\mu \partial_{C'} =
   V^\mu (\partial^{A'} \cA^{B'}_\mu) (\partial_{A'}\otimes\partial_{B'} - \partial_{B'}\otimes\partial_{A'}), 
\end{equation}
and so indeed the absence of such terms in the Lie derivative implies $\partial_{[A'} \cA_{B']}=0$. The absence of the $\partial_{A'}\otimes \hat{\partial}_{B'}$ terms directly implies $\partial_{A'} \hat{\cA}_{B'}=0$, which is equivalent to $\hat{\partial}_{A'} \cA_{B'}=0$. Thus \eqref{HS-SD_GR Twistorspace: Potential as HS field} is a necessary and sufficient condition for the horizontal vector field to be a Poisson symmetry.
\qed

We are thus led to consider Ehresmann connections $\cA(x,\pi)^{A'}$ given by a derivative of the higher-spin field \eqref{HS-SD_GR Spacetime: Potential} (just as for the Yang-Mills case, this is really making use of Hartogs's extension theorem which allows to extend the holomorphicity from $(\mathbb{C}^{2})^*$ to the whole of $\mathbb{C}^{2}$).
 Explicitly, in terms of the higher-spin fields
\begin{equation}
\cA(x,\pi)^{A'} = 
\sum_{n=2}^{\infty} \frac{1}{(n-1)!} A(x)^{A'}{}^{A'(n-1)} \pi_{A'_1}\ldots \pi_{A'_{(n-1)}}.
\end{equation}

In particular, restricting oneself to the case of linear connections (and thus to a connection in the more usual sense of principal bundles),
\begin{equation*}
\gt^{A'}\parD_{A'} = \left(d\pi^{A'} + A(x)^{A'B'}\pi_{B'} \right)\frac{\parD}{\parD \pi^{A'}}
\end{equation*}
 amounts to going from higher spins to the spin-2 field.

At every point $p \in \cT$, curvature of this connection is given by the 2-form
\begin{equation*}
\cF_p \left| 
\begin{array}{ccc}
H_p \times H_p & \to & V_p \\
(X,Y) & \mapsto & P\left([X,Y]\right) 
\end{array}\right.
\end{equation*}
where $[,]$ is the usual Lie Bracket on vector fields, and $P$ is the above projector on the vertical vector fields. 
This definition makes manifest the fact that curvature of a connection is the obstruction to the integrability of the corresponding horizontal distribution. 

It is clear that the required projection on the vertical distribution can be computed as 
\begin{align}\label{HS-SD_GR Twistorspace: Curvature}
-\gt^{0'}\W \gt^{1'} \W \cF^{A'} &= -\gt^{0'}\W \gt^{1'} \W d \gt^{A'}
=  \gt^{0'}\W \gt^{1'} \W \left(  \partial^{A'} d_x A + d\pi^{B'} \partial_{B'} \partial^{A'} A  \right) \\
&= \gt^{0'}\W \gt^{1'} \W \left(  \partial^{A'} d_x A + (\tau^{B'}+\partial^{B'} A) \partial_{B'} \partial^{A'} A\right)
\nonumber\\
 &=  \gt^{0'}\W \gt^{1'}  \left(  \partial^{A'} d_x A +\partial^{B'} A \,\partial^{A'} \partial_{B'}  A \right)= \gt^{0'}\W \gt^{1'}  \partial^{A'}\left( d_x A + \frac{1}{2}  \partial^{B'} A \, \partial_{B'}  A \right)
 \nonumber \\
 &=\gt^{0'}\W \gt^{1'}  \partial^{A'}F.
 \nonumber
\end{align}
Here we denoted by $d_x$ the exterior derivative with respect to the coordinates on the base. We thus see the appearance of the field strength \eqref{HS-SD_GR Spacetime: Field strength}. Here, to pass to the last line we used the fact that $d\pi^{A'}$ can be replaced by $\partial^{A'} A$ on the kernel of $\tau^{A'}$. Thus the ``higher-spin field strength'' \eqref{HS-SD_GR Spacetime: Field strength} genuinely is the curvature of the ``higher-spin potential'' \eqref{HS-SD_GR Spacetime: Potential} when the later is properly understood as an Ehresmann connection $\eqref{HS-SD_GR Twistorspace: Potential}$.

\subsubsection{Field equations}

The field equations \eqref{HS-SD_GR Spacetime: Equations} imply that the 2-form $F$ is decomposable, and thus of the form $F= \gth^{0}\W \gth^{1}$. We now define the following 4-form 
\begin{align}\label{HS-SD_GR Twistorspace: Complex structure}
\gO &:= \gt^{C'}\W \gt_{C'} \W F.
\end{align}
This form is factorisable 
\begin{equation}\label{HS-SD_GR Twistorspace: factorisation of Omega}
\gO= 2\gt^{0'}\W\gt^{1'}\W\gth^{0}\W\gth^{1}.
\end{equation}
We can now define an almost complex structure by requiring $\left(\gt^{0'}, \gt^{1'}, \gth^{0}, \gth^{1}\right)$ to be the basis of $(1,0)$-forms. Thus, we define
$T^*_{(1,0)}\cT := Span\left(\gt^{0'}, \gt^{1'}, \gth^{0}, \gth^{1}\right)$.  The next step is to see whether the HS-SDGR field equations make the almost complex structures defined in this way integrable.

Let $J$ be an almost complex structure, let $\left(\gth^i\right)_{i\in \{1... 4\}}$ be a basis of $(1,0)$-forms and let $\gO$ be the $(4,0)$-form given by $\gO= \gth^{1}\W\gth^{2}\W\gth^{3}\W\gth^{4}$. The Nijenhuis tensor is defined as $\cN^{i} := d\gth^{i} \big|_{(0,2)}$. It can be computed as
\begin{equation*}
\gth^{i} \W d\gO = \cN^{i} \W \gO\,,
\end{equation*}
however, we shall evaluate the components of this tensor directly, in order to better understand the meaning of the equations arising.

Let us start by computing $d\tau^{A'}$. We have
\begin{eqnarray}\label{d-tau}
    d\tau^{A'} = d( d\pi^{A'}-\partial^{A'} A) = d_x \partial^{A'} A + d\pi^{B'}\wedge \partial_{B'}\partial^{A'} A = \\ \nonumber
    = \partial^{A'} d_x A +( \tau^{B'}+\partial^{B'} A)\wedge \partial_{B'}\partial^{A'} A=  \partial^{A'} F +\tau^{B'}\wedge \partial_{B'}\partial^{A'} A.
\end{eqnarray}
A piece of the Nijenhuis tensor of the almost complex structure is obtained by taking the $(0,2)$ component in $d\tau^{A'}$. The last term in (\ref{d-tau}) has a $(1,0)$ $\tau^{B'}$ factor, and so does not survive the projection to $(0,2)$. Therefore, a necessary condition for the almost complex structure to be integrable is
\begin{eqnarray}\label{Nij-1}
    \partial^{A'} F\Big|_{(0,2)}=0.
\end{eqnarray}
This equation indeed follows by differentiating \eqref{HS-SD_GR Spacetime: Equations}. Indeed, we get
\begin{eqnarray}\label{Nij-2}
   (\partial^{A'} F) \W F=0,
\end{eqnarray}
which is equivalent to \eqref{Nij-1}: From the definition \eqref{HS-SD_GR Twistorspace: Complex structure} of the almost complex structure, equation \eqref{Nij-1} is equivalent to $\gt^{0'}\W \gt^{1'} \W F \W  (\partial^{A'} F) =0$ but since $\partial^{A'} F$ is ``horizontal'' (as can be seen e.g. from \eqref{HS-SD_GR Twistorspace: Curvature}) this is equivalent to \eqref{Nij-2}. Thus, there is no obstruction to integrability from $d\tau^{A'}$.

To compute the other components of the Nijenhuis tensor we need to compute the exterior derivative of the decomposable 2-form $F$. We have
\begin{eqnarray}
   dF = d_x F + d\pi^{B'} \partial_{B'} F.
\end{eqnarray}
The first term here can be simplified
\begin{eqnarray}
   d_x F= d_x( d_x A+ \frac{1}{2}\{ A,A\})
   = \{ d_x A,A\} = \{ F, A\},
\end{eqnarray}
which is effectively the Bianchi identity for $F$. The second term can be transformed by replacing $d\pi^{B'}=\tau^{B'} +\partial^{B'} A$. We get
\begin{align}\label{HS-SD_GR Twistorspace: identity1}
   dF &= \{F,A\} + (\tau^{B'} +\partial^{B'} A) \partial_{B'}F \nonumber\\
   &=\{ F,A\} + \{ A,F\} + \tau^{B'} \partial_{B'}F = \tau^{B'} \partial_{B'}F
\end{align}
We want to show that there is no $(1,2)$ component here. However, given that $\tau^{B'}$ is $(1,0)$, this is equivalent to showing that $\partial^{B'} F$ does not have the $(0,2)$ part. As we know from the discussion following  \eqref{Nij-1}, this follows from the field equations by differentiation with respect to $\pi$. Thus, the almost complex structure is integrable, in exact parallel with the spin-2 case. 

\subsubsection{Gauge transformations}
As we will see, the invariance of the HS-SDGR equations under ``higher-spin gauge transformations'' \eqref{HS-SD_GR Spacetime: gauge transformations} and ``generalised diffeomorphisms'' \eqref{HS-SD_GR Spacetime: generalised diffeomorphisms} is essentially equivalent to their invariance under diffeomorphisms of the twistor space.

\paragraph{Higher-spin gauge transformations infinitesimally}
Let $\xi = \xi^{A'}(x,\pi,\pih)\parD_{A'} + \xih^{A'}(x,\pi,\pih)\parDh_{A'}$ be a real vector field along the fibres of $\cT$. The action of the Lie derivative on the Ehresmann connection is 
\begin{align}
\LieD_{\xi}\left(  \gt^{A'}\parD_{A'} \right) &= (d\xi^{A'} -\xi^{B'}\partial_{B'} \partial^{A'} A)\partial_{A'} - \tau^{A'} [\xi,\partial_{A'}]\\
&=
\left( d_x\xi^{A'} +  d\pih^{C'}\parDh_{C'}\xi^{A'}  +\partial^{B'} A \parD_{B'}\xi^{A'}- \xi^{B'} \parD_{B'}\partial^{A'} A \right) \;\parD_{A'} + \left(\tau^{A'}\parD_{A'}\xih^{C'} \right) \;\parDh_{C'}.
\end{align}
Now, $\xi$ is a symmetry of the Poisson structure if and only if
\begin{eqnarray}
   \xi^{A'}(x,\pi,\pih) = \partial^{A'} \xi(x,\pi)
\end{eqnarray}
i.e. if and only if it is hamiltonian and holomorphic. We get in this case
\begin{align}\nonumber
\LieD_{\xi}\left(  \gt^{A'}\parD_{A'} \right) =
\left( d_x\partial^{A'} \xi +\partial^{B'} A \parD_{B'}\partial^{A'}\xi- \partial^{B'}\xi \parD_{B'}\partial^{A'} A  \right) \;\parD_{A'}
= \partial^{A'} (d_x \xi+\{A,\xi\}) \;\parD_{A'}.
\end{align}
This last expression corresponds to the ``higher-spin gauge transformations'' \eqref{HS-SD_GR Spacetime: gauge transformations} which therefore correspond to infinitesimal vertical (Poisson) diffeomorphisms in twistor space.

\paragraph{Non-linear realisation of higher-spin gauge transformations}

As we just saw, higher-spin gauge transformations can be interpreted as the action of the Lie derivative along the fibres. This suggests to consider the non-linear action of vertical automorphisms of the bundle 
\begin{equation*}
    f \left|\begin{array}{ccc}
\mathbb{T} &\to& \mathbb{T} \\
  (x^{\mu},\pi_{A'}) &\mapsto& ( x^{\mu} , f_{A'}(x,\pi) )
    \end{array}\right.
\end{equation*}
which are Poisson symmetries 
\begin{equation}
f_*\left( \epsilon^{A'B'}\parD_{A'}\parD_{B'} + \epsilon^{A'B'}\parDh_{A'}\parDh_{B'} \right) = \epsilon^{A'B'}\parD_{A'}\parD_{B'} + \epsilon^{A'B'}\parDh_{A'}\parDh_{B'} 
\end{equation}

We now want to show that these vertical Poisson diffeomorphisms act on the space of connections \eqref{HS-SD_GR Twistorspace: Potential} satisfying \eqref{HS-SD_GR Twistorspace: Potential as HS field} and thus provide a non-linear realisation of the "higher-spin gauge transformations".

In order to prove this let us consider the horizontal distribution $D_{\mu}$ given by \eqref{HS-SD_GR Twistorspace: Horizontal vector field} and its push-forward $f_{*}(D_{\mu})$ under a vertical Poisson diffeomorphism. As we know from \eqref{prop-horizontal}, $f_{*}(D_{\mu})$ defines a connection satisfying \eqref{HS-SD_GR Twistorspace: Potential as HS field} if and only if
\begin{equation*}
[f_{*}\big(D_{\mu}\big) , \epsilon^{A'B'}\parD_{A'}\parD_{B'} + \epsilon^{A'B'}\parDh_{A'}\parDh_{B'}  ] =0\,.
\end{equation*}
We then have
\begin{align}\notag
[f_{*}\big(D_{\mu}\big) , \epsilon^{A'B'}\parD_{A'}\parD_{B'} + \epsilon^{A'B'}\parDh_{A'}\parDh_{B'}  ] &= [f_{*}\big(D_{\mu}\big) , f_{*}\big(\epsilon^{A'B'}\parD_{A'}\parD_{B'} + \epsilon^{A'B'}\parDh_{A'}\parDh_{B'}  \big)]\\
&= f_{*}\big([D_{\mu} , \epsilon^{A'B'}\parD_{A'}\parD_{B'} + \epsilon^{A'B'}\parDh_{A'}\parDh_{B'}  ]\big)\\
&=0\,,\notag
\end{align}
where the first equality uses that $f$ is a Poisson symmetry, the second follows from the identity $f_* [X,Y] = [f_* X , f_* Y]$ and the third from our assumptions on $D_{\mu}$. Thus, indeed vertical Poisson diffeomorphisms act on the space of connections satisfying \eqref{HS-SD_GR Twistorspace: Potential as HS field} and provide the non-linear realisation of the higher-spin gauge symmetry. 

In order to see the action of these diffeomorphisms on the field equations we need to see how it acts on the curvature. However $F^{A'} =\parD^{A'}F = \LieD_{\parD_{A'}}F$ is the curvature of $\tau^{A'}$ and thus a tensorial object. It follows that $F$ is simply a 2-form.

The invariance of the field equations under the higher-spin gauge transformations directly follows from their invariance under the vertical Poisson diffeomorphisms.

\paragraph{Generalised diffeomorphisms} The situation with generalised diffeomorphisms is more subtle, as it is not easy to provide them with a clear geometrical interpretation. In fact, we shall see that these transformations are in general not diffeomorphisms, and so it would be better to call the symmetry of the field equations that they generate "the shift symmetry" rather than "generalised diffeomorphisms".

We start by verifying that these transformations are indeed a symmetry of the field equations. Let us consider horizontal vector fields of the form $\eta = \eta\left(x,\pi\right)^{\mu}\parD_{\mu}$ and the shift symmetry
\begin{equation}
    \delta A = \eta \intD F\,.
\end{equation}
In order to prove the (on-shell) invariance of the field equation
\begin{equation}
\delta_{\eta}\left( F \wedge F\right) = 2 \delta_{\eta} F \wedge F= 0
\end{equation}
we will need to rewrite the variation as
\begin{align}
\delta_{\eta} F & = d_x(\eta \intD F) + \{\eta \intD F, A\} \notag \\ \nonumber
& = d(\eta \intD F)- d\pi^{A'} \partial_{A'} ( \eta\intD F) + \partial^{A'} (\eta \intD F) \partial_{A'}A \\ 
&= d(\eta \intD F)- \tau^{A'} \LieD_{\parD_{A'}} ( \eta\intD F) \\ \nonumber
&= \LieD_{\eta}F - \eta \intD dF - \tau^{A'}\big( (\LieD_{\parD_{A'}}\eta)\intD F) + (\eta \intD (\LieD_{\parD_{A'}} F)\big)\\ &= \LieD_{\eta}F - \tau^{A'} (\LieD_{\parD_{A'}}\eta)\intD F + \eta\intD \left( -dF + \tau^{A'}\parD_{A'}F\right)\nonumber\\
&= \LieD_{\eta}F - \tau^{A'} (\LieD_{\parD_{A'}}\eta)\intD F\notag
\end{align}
where to get to the line before last we exchanged the insertion of $\eta$ with $\tau^{A'}$ and thus got an additional minus sign, and to get to the last line we used the identity \eqref{HS-SD_GR Twistorspace: identity1}. It follows that
\begin{align} \delta_{\eta} F \wedge F &= \left(\LieD_{\eta}F - \tau^{A'} (\LieD_{\parD_{A'}}\eta)\intD F\right)\wedge F \notag \\
&= \frac{1}{2}\LieD_{\eta}\left(F \wedge F \right) -\frac{1}{2} \tau^{A'}(\LieD_{\parD_{A'}}\eta)\intD \left(F\wedge F \right)\\
&=0\,.\notag
\end{align}
Where in the last step we used the field equation.

However, as noted already in \cite{Krasnov:2021nsq}, generalised diffeomorphisms do not coincide with the Lie derivative in the direction of the vector field $\eta$. The disagreement is by terms containing the derivative of $\eta^\mu(x,\pi)$ with respect to $\pi$. Indeed, we consider the vector field $\eta^\mu D_\mu= \eta^\mu(\partial_\mu + \partial^{A'} A_\mu \partial_{A'})$. We have
\begin{align}\nonumber
\LieD_{(\eta^\mu D_\mu)}\left( \gt ^{A'}\parD_{A'}  \right)  &=
-\eta^\mu (\partial_\mu \partial^{B'} A)\partial_{B'} - \eta^\mu (\partial^{B'} A_\mu) (\partial_{B'} \partial^{A'} A) \partial_{A'} 
\\ \label{eta-Lie}
& + d(\eta^\mu \partial^{A'} A_\mu) \partial_{A'} - (d\eta^\mu)(\partial^{A'}A_\mu) \partial_{A'} + \tau^{A'} [\eta^\mu D_\mu,\partial_{A'}] \\\nonumber
&= - (\eta\intD \partial^{A'} F) \partial_{A'} - \tau^{A'} (\partial_{A'}\eta^\mu) D_\mu\, \\ \nonumber
&= - \partial^{A'} (\eta\intD F) \partial_{A'} + (\partial^{A'}\eta)\intD F \partial_{A'}- \tau^{A'} (\partial_{A'}\eta^\mu) D_\mu\,,
\end{align}
where we have used
\begin{eqnarray}
   (\eta\intD \partial^{A'} F)_\nu = \eta^\mu \partial_\mu \partial^{A'} A_\nu - \eta^\mu \partial_\nu \partial^{A'} A_\mu + \eta^\mu \partial^{B'} \partial^{A'} A_\mu \partial_{B'}A_\nu - \eta^\mu \partial^{B'} \partial^{A'} A_\nu \partial_{B'}A_\mu\,.
\end{eqnarray}
The first term in \eqref{eta-Lie} is the desired generalised diffeomorphism 
\begin{equation*}
    \delta_\eta (\tau^{A'} \partial_{A'}) = - \partial^{A'}(\eta\intD F) \partial_{A'}\,.
\end{equation*}
The remaining terms all contain $\partial_{A'} \eta^\mu$. So, in general the generalised diffeomorphism does not coincide with the Lie derivative by terms containing $\partial_{A'} \eta^\mu$. A similar type of an obstruction appeared in \cite{Krasnov:2021nsq}, cf. (3.39) in \cite{Krasnov:2021nsq} with \eqref{eta-Lie}. Therefore, as we already anticipated, the generalised diffeomorphisms represent a gauge symmetry of the equations, but they cannot in general be identified with genuine diffeomorphisms.

\subsection{A higher-spin non-linear graviton correspondence}

The discussion above can be summarised as the following theorem.

Let $S' \xto{\pi} M^4$ be the bundle of spinors on $M^4$, the twistor space $\cT \xto{\pi} M^4 $ is obtained by deleting from $S'$ the zero section.  Let $U$ be an open set of $M^4$ and let $V = \pi^*(U)$ be the corresponding open set of $\cT$. We define the horizontal distributions on $TV$ to be those in the kernel of the projection $P \from T\cT \to \mathcal{V}$
	\begin{equation*}
	P =  \gt^{A'}\frac{\parD}{\parD \pi^{A'}} +  \gthat^{A'}\frac{\parD}{\parD \pih^{A'}}, \qquad \tau^{A'}:= d\pi^{A'} + \cA^{A'}(x,\pi,\hat{\pi})\,.
	\end{equation*}
They are parametrised by Ehresmann connection $\cA^{A'}(x,\pi,\hat{\pi})$.
\begin{Theorem}
For horizontal distributions 
that are infinitesimal symmetries of the Poisson structure \eqref{HS-SD_GR Twistorspace: Poisson structure} the Ehresmann connection $\cA^{A'}$ is potential: $\cA^{A'}=-\epsilon^{A'B'} \partial_{B'} A$, where $A=A(x,\pi)$. In particular, the Ehresmann connection $\cA^{A'}$ of Poisson horizontal distributions is independent of $\hat{\pi}$. Furthermore, its curvature 2-form is also potential $\cF^{A'}=-\epsilon^{A'B'}\partial_{B'} F$, where $F=F(x,\pi)=dA+(1/2)\{A,A\}$. There is a one-to-one correspondence (up to a gauge) between solutions of the higher-spin self-dual gravity equations on $U$, with $A(x,\pi)$ as the generating function \eqref{HS-SD_GR Spacetime: Potential}, and Poisson horizontal distributions on $TV$ that have decomposable $F\W F=0$ curvature potential $F$. Together with $\tau^{A'}$, the two simple factors of $F$ define an almost complex structure on $V$ that is integrable. 
\end{Theorem}

\section*{Acknowledgments}
\label{sec:Aknowledgements}
Y.H. and E.S. are glad to acknowledge scientific exchanges with Tim Adamo. The work of Y.H. and E.S. was partially supported by the European Research Council (ERC) under the European Union’s Horizon 2020 research and innovation programme (grant agreement No 101002551). The authors are grateful to Lionel Mason for a conversation that led to a correction in the statement of Theorem 2.1.

\footnotesize
\bibliographystyle{utphys}
\bibliography{megabib}

\end{document}